\documentclass[aps,prb,twocolumn,superscriptaddress]{revtex4}
\usepackage{graphicx}
\usepackage{color}
\usepackage{amssymb}
\usepackage{amsmath}

\usepackage{hyperref}

\newcommand{\supiso}{^{({\rm iso})}}
\newcommand{\bI}{{\bf I}}

\newcommand{\bJ}{{\bf J}}
\newcommand{\bB}{{\bf b}}
\newcommand{\ket}[1]{|#1\rangle}
\newcommand{\bra}[1]{\langle #1|}

\newcommand{\rhoin}{\rho}
\newcommand{\omNOON}{{\omega_{\rm \noon}}}
\newcommand{\diag}{diag}

\newcommand{\lpp}{s_{++}}
\newcommand{\lpm}{s_{+-}}
\newcommand{\lmp}{s_{-+}}
\newcommand{\lmm}{s_{--}}
\newcommand{\nmeas}{M}

\newcommand{\noon}{{NOON}~}
\newcommand{\scat}{{\cal S}}
\newcommand{\cespdc}{{CESPDC}}

\begin{document}

\title{Entanglement-enhanced probing of a delicate material system}

\author{Florian Wolfgramm}
\affiliation{ICFO -- Institut de Ciencies Fotoniques, Mediterranean Technology Park, 08860 Castelldefels (Barcelona), Spain}

\author{Chiara Vitelli}
\affiliation{Center of Life Nanoscience at La Sapienza, Istituto Italiano di Tecnologia, Viale Regina Elena 255, 00181 Rome, Italy}

\author{Federica A. Beduini}
\affiliation{ICFO -- Institut de Ciencies Fotoniques, Mediterranean Technology Park, 08860 Castelldefels (Barcelona), Spain}

\author{Nicolas Godbout}
\affiliation{COPL, D\'epartement de G\'enie Physique, \'Ecole Polytechnique de Montr\'eal, C.P. 6079, Succ. Centre-ville, Montr\'eal (Qu\'ebec) H3C 3A7, Canada}

\author{Morgan W. Mitchell}
\affiliation{ICFO -- Institut de Ciencies Fotoniques, Mediterranean Technology Park, 08860 Castelldefels (Barcelona), Spain}
\affiliation{ICREA -- Instituci\'{o} Catalana de Recerca i Estudis Avan\c{c}ats, 08015 Barcelona, Spain}

\maketitle

{\bf Quantum metrology \cite{Giovannetti2011} uses entanglement \cite{Leibfried2004, Mitchell2004, Vahlbruch2008, Afek2010} and other quantum effects \cite{Napolitano2011} to improve the sensitivity of demanding measurements \cite{Wolfgramm2010a,Shah2010,LIGO2011}. Probing of delicate systems demands high sensitivity from limited probe energy and motivates the field's key benchmark the standard quantum limit \cite{Braunstein2006}. Here we report the first quantum-enhanced measurement of a delicate material system. We non-destructively probe an atomic spin ensemble by near-resonant Faraday rotation, a measurement that is limited by probe-induced scattering in quantum memory and spin-squeezing applications \cite{Madsen2004,Julsgaard2004,Koschorreck2010b,Napolitano2011}. We use narrowband, atom-resonant \noon states to beat the standard quantum limit of sensitivity by more than five standard deviations, both on a per-photon and a per-damage basis. This demonstrates quantum enhancement with fully realistic loss and noise, including variable-loss effects\cite{Kacprowicz2010,Thomas-Peter2011,Escher2011}. The experiment points the way to ultra-gentle probing of single atoms \cite{Tey2008}, single molecules \cite{Pototschnig2011}, quantum gases \cite{Eckert2008} and living cells \cite{Carlton2010}.}

Linear interferometry with non-entangled states can reach at best the standard quantum limit (SQL) $\delta \phi = 1/\sqrt{N}$, where $\phi$ is the interferometric phase to be measured and $N$ is the number of probe particles. When increasing $N$ is not possible, quantum enhancement offers a practical advantage. A key example is interferometric gravitational-wave detection: in current operating conditions, squeezing improves sensitivity whereas increasing photon flux produces deleterious thermal effects \cite{LIGO2011}. Here we study an analogous number-limited scenario with very broad potential application: the probing of delicate systems, i.e., material systems which suffer significant damage due to the probing process. Examples are found in atomic \cite{Tey2008}, molecular \cite{Pototschnig2011}, condensed matter \cite{Eckert2008} and biological \cite{Carlton2010} science.

By the Kramers-Kronig relations, interferometric phase shifts are necessarily accompanied by absorption, implying deposition of energy in the probed medium. Absorption also degrades any quantum advantage, as described by recent theory \cite{Dorner2009,Banaszek2009,Escher2011}. To further complicate matters, in real media the phase shift and absorption may depend on the same unknown quantity. To show advantage in a fully-realistic scenario, we probe a precisely understood material system using a quantum state permitting rigorous sensitivity and damage analysis.

Our delicate system is a $^{85}$Rb atomic spin ensemble, similar to ensembles used for optical quantum memories \cite{Julsgaard2004} and quantum-enhanced atom interferometry
\cite{Schleier-Smith2010}. Non-destructive dispersive measurements on these systems, used for storage and readout of quantum information or to produce spin squeezing,
are fundamentally limited by scattering-induced depolarization \cite{Madsen2004,Julsgaard2004,Koschorreck2010b,Napolitano2011}.
Both the measurement and damage properties of the ensemble can be calculated from first principles, making this an ideal model system. 

We probe the ensemble with a polarization \noon state, a two-mode entangled state of the form $|N\rangle_L|0\rangle_R + \exp[i \phi] |0\rangle_L|N\rangle_R$ (normalization omitted), in which $N$ particles are either all $L$- or all $R$-circularly 
polarized. The use of entangled photons in the single-photon regime allows a rigorous quantification, by quantum state tomography and Fisher Information (FI) theory, of the information gained. To minimize losses while maximizing rotation, we employ a \noon state in a 7 MHz spectral window detuned four Doppler widths from the nearest $^{85}$Rb resonance. This requires matter-resonant indistinguishable photons, which we generate for the first time with an advanced down-conversion source and ultra-narrow atom-based filter.

We study the ratio of FI to probe-induced damage, i.e., to photons absorbed (equivalently scattered) by the ensemble. Since both FI and damage scale linearly with the number of probe particles, the same advantage is gained for higher photon numbers, e.g., at the projection-noise level. We find the \noon state beats the SQL by $30 \pm 5$\% in information gained per photon and $23 \pm 4$\% per damage to the ensemble.

\begin{figure}[t]
\centering
\includegraphics[width=0.45\textwidth]{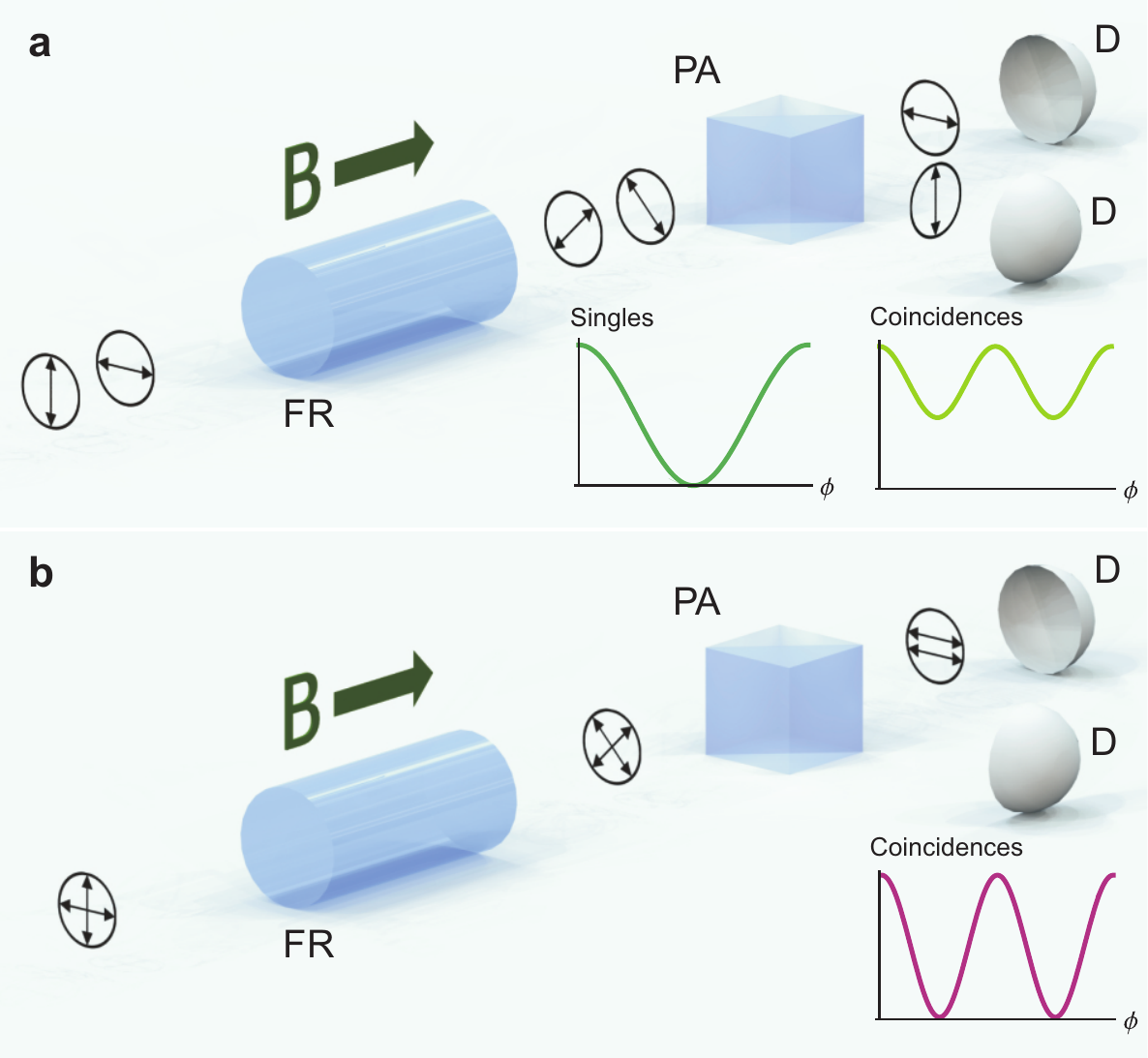}
\caption{\noon state probing of an \ {atomic 
ensemble by Faraday rotation}. Polarized photons pass through a rubidium 
vapor cell (FR) where they
experience an optical rotation dependent on the applied magnetic
field $\vec{B}$ (the Faraday effect) and are detected by a
polarization analyzer (PA) with photon counting detectors (D).
{\bf a} With non-entangled photons, ``singles detection'' can give
high visibility, while ``coincidence detection'' can give (low
visibility) super-resolution. {\bf b} By Hong-Ou-Mandel
interference, \noon states can give both super-resolution and high
visibility, providing more information per photon than possible
without entanglement. In repeated probing, fewer photons are
needed to achieve the same sensitivity, causing less damage to the
sample.
\label{fig:Cell}
}
\end{figure}

The setup is shown schematically in Fig. \ref{fig:Cell} {\bf b}: narrowband \noon states at $\omNOON$, the 
optical frequency of the $5^{2}{\rm S}_{1/2} {\rm F}$=$2 \rightarrow 5^{2}{\rm P}_{1/2} {\rm F}'$=$1$ transition of the $D_1$ line of $^{87}$Rb, are generated by cavity-enhanced spontaneous parametric down-conversion (\cespdc) and sent through the ensemble. The ensemble of $^{85}$Rb atoms is contained in an anti-reflection coated vapor cell with internal length $L=75$ mm, in a temperature-controlled oven at 70$^\circ$ C, together with a 0.5\% residual $^{87}$Rb component. An applied axial magnetic field $B$ of up to 60 mT produces resonantly-enhanced Faraday rotation of the optical polarization. After leaving the vapor cell, the photons are separated in polarization, frequency filtered, and
detected with single-photon counters. Two counters on each polarization output record all possible outcomes, i.e., coincidences of $HH$, $HV$, and $VV$ polarizations are post-selected.

\begin{figure}[t]
\centering
\includegraphics[width=0.45\textwidth]{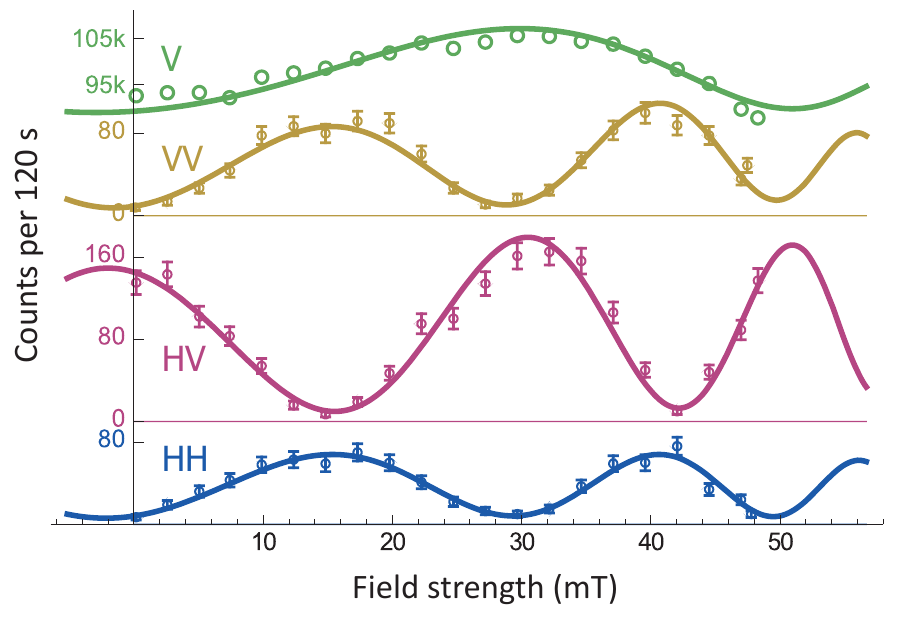}
\caption{High-visibility super-resolving Faraday rotation probing using optical \noon states. Top curve: for phase reference, singles detection rate (V polarization) {versus field strength B} 
shows one oscillation in the range B=0 to 50 mT. Other curves: coincidence detections $HH$, $HV$, and $VV$ show two oscillations in the same range (super-resolution) and high visibility. Symbols show measured data (no background subtracted), with $\pm 1 \sigma$ statistical uncertainties. Curves show model of Eq. (\ref{eq:CoincProbabilities}) with state $\rhoin$ found by quantum state tomography.
\label{fig:SuperRes}}
\end{figure}

As seen in Fig. \ref{fig:SuperRes}, all coincidence outcomes oscillate as a function of $B$, with two-fold 
super-resolution relative to the single-photon oscillation, visible in the singles counts due to a small imbalance between $H$ and $V$ in the input state. The interference visibilities are all $\ge90$\%, well above the 33\% classical limit for $HH$ and $VV$ visibility\cite{Afek2010a}. Also shown are predicted coincidence rates $R_i = R_0 P_i$, where $R_0$ is the input flux of pairs and 
\begin{equation}
P_i(B) = {\rm Tr}[\Pi_i
T(B) \rhoin T^\dagger(B) ],
\label{eq:CoincProbabilities}
\end{equation}
are the outcome probabilities. Here $\rhoin$ is the two-photon state before the cell, $B$ is the magnetic field, $T(B)$ describes the transmission through the cell, and $\Pi_i$ is the positive-operator-valued measure (POVM) element for the $i$'th outcome. 

In the $\sigma_\pm$ basis, $T(B) = {\rm \diag}(t_+^2,t_+t_-,t_-t_+,t_-^2)$, where $t_\pm = \exp[i n_\pm \omega_{\rm \noon} L / c]$ 
are transmission coefficients and $n_\pm(B)$ are the complex refractive indices from a first-principles calculation (see supplementary information). In the H,V basis, $\Pi_{HH},\Pi_{HV}$ and $\Pi_{VV}$ are $\rm{\diag}(1,0,0,0)$, $\rm{\diag}(0,1,1,0)$, and $\rm{\diag}(0,0,0,1)$, respectively. A completely analogous
description is used for single photon probabilities. Due to atomic absorption and scattering, $|t_{\pm}| < 1$ so that $\sum_i P_i < 1$ in general.

\begin{figure}[b]
\centering
\includegraphics[width=0.45\textwidth]{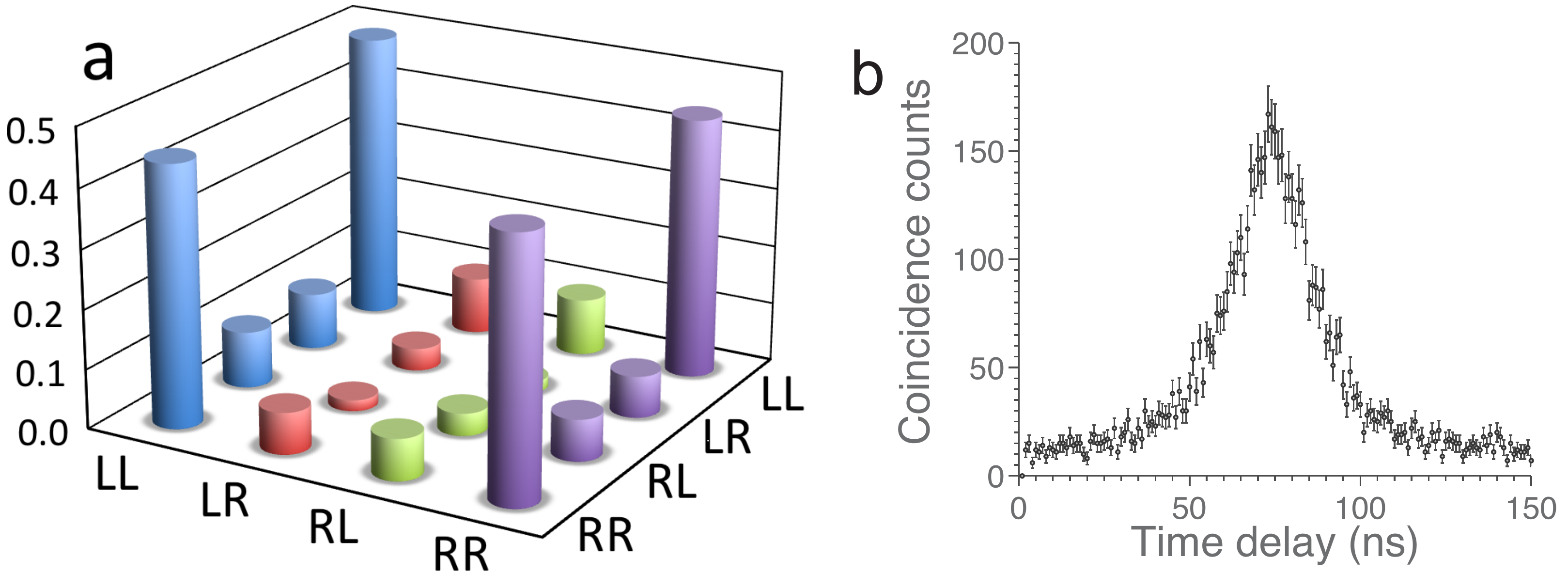}
\caption{\noon state characterization. {\bf a} Density matrix $\rhoin$ (magnitudes only)
 from quantum state tomography, showing large coherence between $\ket{LL}$ and
$\ket{RR}$ components. {\bf b} Measured correlation of the filtered \cespdc~pairs (no background subtracted). The absence of modulation at the 2 ns cavity round-trip time indicates the presence of a single cavity mode.
\label{fig:DM_TC}}
\end{figure}

The $\rho$ in Eq. (\ref{eq:CoincProbabilities}) is found by quantum state tomography \cite{Adamson2007}, i.e., a fit to the observed data using Eq. (\ref{eq:CoincProbabilities}) and the known $T(B)$ and $\Pi_i$. We find a state, shown in {Fig. \ref{fig:DM_TC}.{\bf a}}, with high purity ${\rm Tr}[\rho^2] = 0.88$, low 
photon distinguishability \cite{Adamson2007} $\bra{\psi^-} \rhoin \ket{\psi^-} = 0.02$, and high fidelity $\bra{{N_\phi}} \rhoin \ket{{N_\phi}} = 0.90$ with the \noon state $\ket{{N_\phi}} \equiv (\ket{2_L 0_R} + \exp[2i \phi] \ket{0_L 2_R}) / \sqrt{2}$, $\phi =0.22$. The phase $\phi$ is twice the input polarization rotation, which can be chosen to optimize sensitivity at any given value of $B$. Note that both rotation and absorption of the \noon photons is evident in Fig. \ref{fig:SuperRes}, and contributes to the state reconstruction. A simultaneously-acquired time-correlation measurement, shown in Fig. \ref{fig:DM_TC} {\bf b} confirms the spectral purity of the \noon state \cite{Kuklewicz2006}. To our knowledge, this is the first demonstration of an atom-resonant state with multi-photon coherence. 

We rigorously quantify the metrological advantage using the Fisher information (FI),
\begin{equation}
{\cal I}(B) \equiv \sum_i P_i |\partial_B \ln P_i|^2,
\end{equation}
where $P_i$ are the probabilities derived from the first-principles model in Eq. (\ref{eq:CoincProbabilities}). The magnetic uncertainty is $\delta B \rightarrow ({\cal I}\nmeas)^{-1/2}$ in the practical scenario of $M \gg 1$ uses of the state \cite{Fisher1925}. On the basis of FI per input photon, the \noon state achieves an advantage ${\cal I}_{\rm \noon} / 2{\cal I}_{\rm SQL}=1.30 \pm 0.05$ over the SQL found by numerical optimization over possible single-photon inputs and the POVM $\{ \Pi_i \}$ (details in supplementary information). Despite the losses inherent in near-resonant probing of the ensemble, the \noon state offers a significant advantage. 

\begin{figure}[t]
\centering
\includegraphics[width=0.45\textwidth]{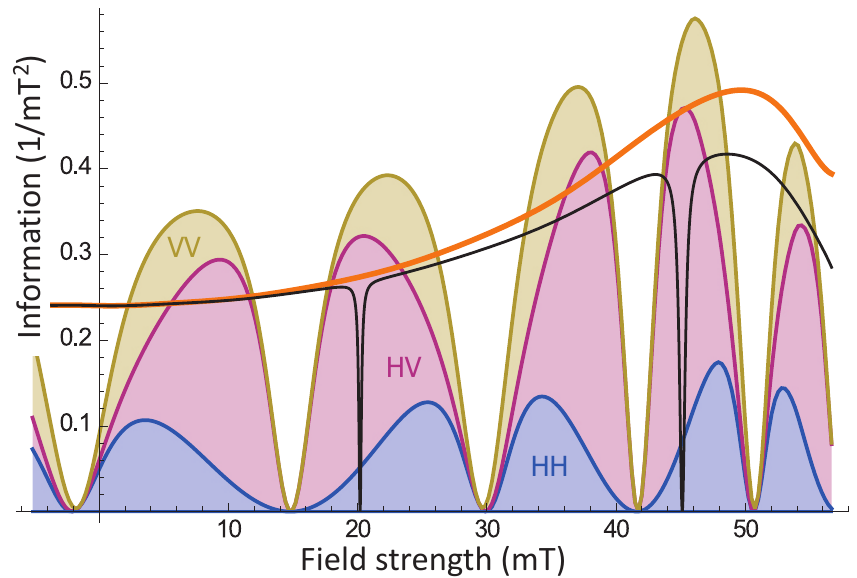}
\caption{Quantum-enhancement in probing of a delicate system, quantified by Fisher information (FI) per mean scattering (S) from the $^{85}$Rb atomic ensemble. Thin black curve: FI/S for a
single photon of an arbitrarily chosen linear input polarization. Thick orange curve: SQL, the largest FI/S obtainable with non-entangled states, found by numerical optimization of a single
photon input and the measurement POVM. Shaded regions: contributions of $HH$, $HV$, and $VV$ outcomes to the FI/S with the experimental \noon state of Fig. \ref{fig:DM_TC}. The \noon state surpasses the SQL by $40 \pm 6 $\% ($23 \pm 4$\% with $^{87}$Rb contaminant), giving a metrological advantage in the presence of fully realistic and parameter-dependent losses. Low points in the FI are associated with extrema of the coincidence probability curves, i.e., regons of small signal sensitivity. For both single-photon and \noon-state probing, the input polarization determines the position of the maxima, allowing high sensitivity measurement for any value of $B$.
\label{fig:FIpdMain}
}
\end{figure}

The \noon state also gives an advantage in FI per {scattered} photon. This figure of merit quantifies the information gain per damage and differs from FI per input photon because the scattering is strongly B-dependent. We quantify scattering from the $^{85}$Rb ensemble as $\scat = {\rm Tr} [\rho \Pi_{\rm scat}]$, where $\Pi_{\rm scat} \equiv {\rm diag}(\lpp,\lpm,\lmp,\lmm)$ in the $\sigma_{\pm}$ basis and the mean number of scattering events is $s_{ab} \equiv 2-|t_a^{(85)}|^2-|t_b^{(85)}|^2$, where $t_\pm^{(85)}$ is the $^{85}$Rb contribution to $t_\pm$. A completely analogous calculation is made for single-photon scattering. The \noon state gives an advantage $({\cal I}/\scat)_{\rm \noon}/({\cal I}/\scat)_{\rm SQL} = 1.23 \pm 0.04$
over the SQL with our ensemble including the $^{87}$Rb contaminant. If this were removed, the advantage would be $1.40 \pm 0.06$. Shown graphically in Fig. \ref{fig:FIpdMain}.

These post-selected results imply also an advantage without post-selection: with available detector efficiency\cite{Lita2008} $\eta_{\rm det} = 0.95$ and source-to-detector path efficiency\cite{Vahlbruch2008} $\eta_{\rm path} = 0.984$ the quantum enhancement is $1.21 \pm 0.05$ per sent photon and $1.15 \pm 0.04$ per photon scattered from the $^{85}$Rb ensemble ($1.31 \pm 0.06$ without $^{87}$Rb). In the future, narrow-band squeezed states \cite{Wolfgramm2010a} may allow larger enhancements\cite{Vahlbruch2008} with bright beams and higher detection efficiencies, although squeezing-induced damage is predicted in some scenarios \cite{Auzinsh2004}. See supplementary information.
We note a previously-unreported source of metrological advantage. In the ensemble, as in any material system, the loss depends on the measured quantity (here $B$). This dependence makes a positive contribution to the FI when it increases $|\partial_B \ln P_i|^2$, offsetting the well-known \cite{Thomas-Peter2011} FI reduction from $\sum_i P_i < 1$. 

We have demonstrated quantum-enhanced probing of an intrinsically delicate system, an atomic spin ensemble. Using narrowband, cavity-enhanced down-conversion, we generate a high-fidelity photonic \noon state tuned to an optical resonance of atomic rubidium. When used to probe a rubidium atomic ensemble, the \noon state surpasses the best possible non-entangled probe. The result shows that quantum entanglement can produce a more gentle measurement without sacrificing sensitivity. This is also to our knowledge the first use of multi-photon coherence in an atomic physics context. Quantum-enhanced probing of atomic ensembles paves the way for ultra-gentle probing of {other} delicate systems including exotic phases of cold quantum gases \cite{Eckert2008}, single molecules \cite{Pototschnig2011} and living cells \cite{Carlton2010}.

\section*{METHODS SUMMARY}
We use type-II \cespdc \cite{Kuklewicz2006, Wolfgramm2008, Wolfgramm2010} to produce $N=2$ \noon states. A 795 nm laser, stabilized to $\omNOON$, the frequency of the $5^{2}{\rm S}_{1/2} {\rm F}$=$2 \rightarrow 5^{2}{\rm P}_{1/2} {\rm F}'$=$1$ transition of the $D_1$ line of $^{87}$Rb, is both used to stabilize the cavity length, and is frequency-doubled to pump the SPDC process. In the down-conversion process, the cavity resonantly enhances degenerate down-conversion into a TEM$_{00}$ mode with bandwidth $7$ MHz around $\omNOON$. Non-degenerate emission, separated by at least the cavity's 490 MHz free spectral range, is efficiently removed by filtering. The apparatus is described in detail in references
\cite{Wolfgramm2008,Wolfgramm2010}. By using a low pump power, we ensure that the emitted photons are in a two-photon state $\rhoin$, ideally the \noon state $|1\rangle_H|1\rangle_V \propto |2\rangle_L|0\rangle_R - |0\rangle_L|2\rangle_R $, with negligible four-photon component. $H,V,L,R$ indicate horizontal, vertical, left circular (or $\sigma_+$) and right circular (or $\sigma_-$) polarizations, respectively.

The filter, described in references \cite{Cere2009,Wolfgramm2011}, has an 80 MHz FWHM passband centered on $\omNOON$ and $>$35 dB out-of-band rejection, so that only atom-tuned photons are detected. The single-mode character is evident in the double-exponential arrival-time distribution of the filtered \noon state, shown in Fig. \ref{fig:DM_TC}.{\bf b}. The combination of \cespdc~and narrowband filtering produces a state 
with ideal characteristics for atomic probing: single-spatial
mode, near-perfect indistinguishability, and extremely high
temporal coherence. In a previous experiment \cite{Wolfgramm2011}, we showed at least 94\% of photon pairs are rubidium resonant.

~ \\ 
\noindent 
{\bf Acknowledgements:}
The authors thank A. Cer\`{e}, Y. A. de Icaza Astiz, M. Napolitano, N. Behbood, R. J. Sewell, and M. Hendrych for useful discussions. This work was supported by the Spanish Ministerio de Economia y Competitividad under the project Magnetometria Optica (No. FIS2011-23520), by the European Research Council starting grant ``AQuMet'', by Fundaci\'o Privada Cellex and by an Institute of Photonic Sciences -- Ontario Centres of Excellence collaborative research programme. F.W. is supported by the Commission for Universities and Research of the Department of Innovation, Universities and Enterprises of the Catalan Government and the European Social Fund. N.G. acknowledges support from the National Sciences and Engineering Research Council of Canada.

~ \\ ~ \\
\noindent 
{\bf Author contributions:}
F.W. and M.W.M. conceived and designed the project. F.W., C.V. and F.A.B. designed, constructed and tested the apparatus. F.W. acquired the data and performed the analysis. M.W.M. and N.G. performed the modelling and simulation. All authors contributed to the preparation of the manuscript.

\section*{SUPPLEMENTARY INFORMATION}

\subsection{Model of atomic vapor magneto-optical properties}
\begin{figure}[b]
\centering
\includegraphics[width=0.45\textwidth]{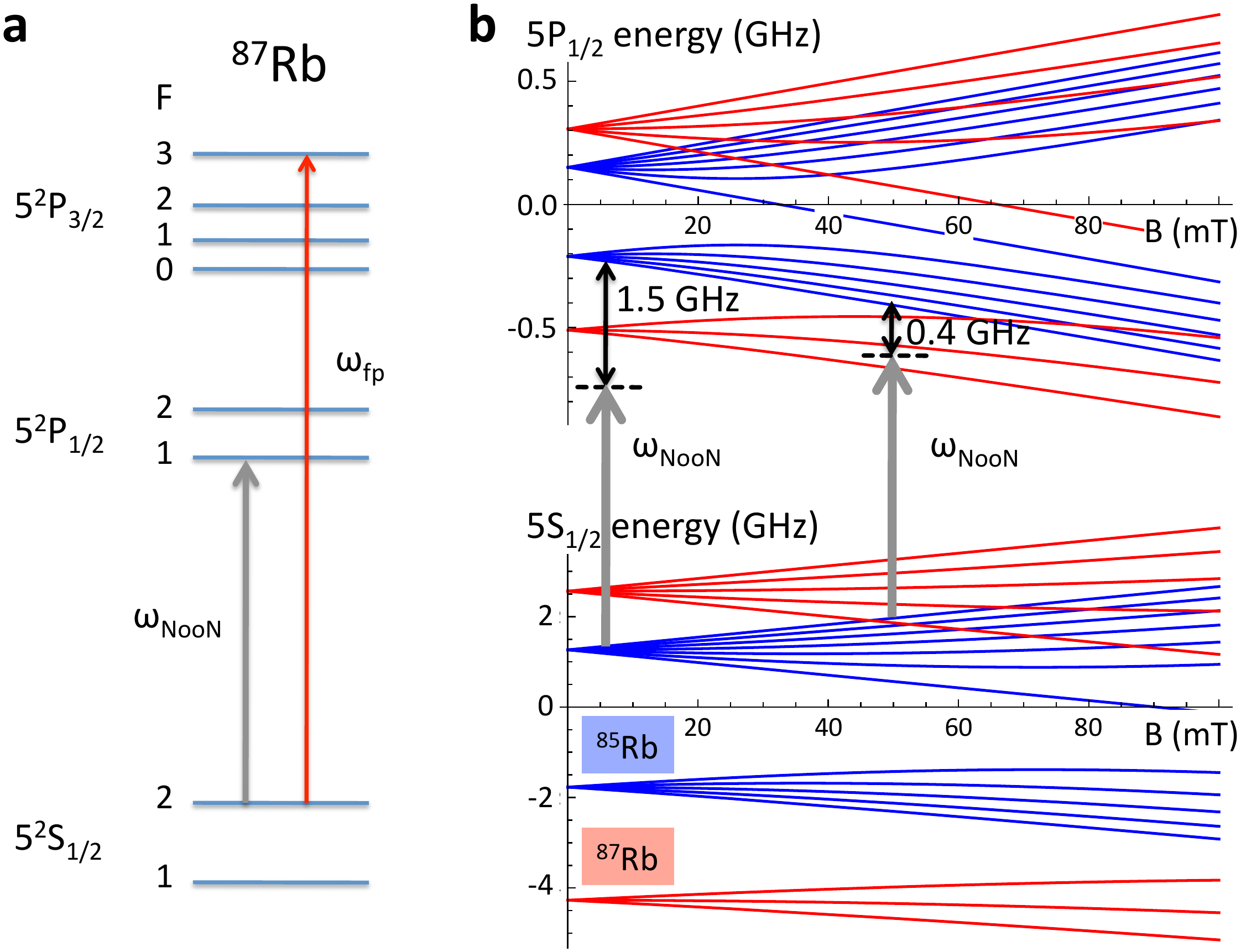}
\caption{Relevant energy level diagrams. {\bf a} Energy levels of $^{87}$Rb relevant to generation and filtering (not to scale). The frequency of the NooN state $\omega_{\rm NooN}$ is tuned to the $5^{2}{\rm S}_{1/2} {\rm F}$=$2 \rightarrow 5^{2}{\rm P}_{1/2} {\rm F}'$=$1$ transition of the $D_1$ line of $^{87}$Rb. The optical pumping laser of the filter, with frequency $\omega_{\rm fp}$, addresses the $5^{2}{\rm S}_{1/2} {\rm F}$=$2 \rightarrow 5^{2}{\rm P}_{3/2} {\rm F}'$=$3$ transition of the $D_2$ line of $^{87}$Rb. The 15 nm separation from the detection wavelength allows a high extinction using interference filters centered on $\omega_{\rm NooN}$. {\bf b} $D_1$ energy levels of the probed ensemble versus field strength $B$, showing $^{85}$Rb levels in blue and $^{87}$Rb levels in red. At zero field $\omega_{\rm \noon}$ is 1.5 GHz detuned from the nearest $^{85}$Rb transition. With increasing $B$, the nearest $^{85}$Rb transition moves closer to resonance, increasing the Faraday rotation. The Doppler-broadened absorption begins to overlap $\omega_{\rm \noon}$ near $B=$ {50} {mT}. 
\label{fig:ED}}
\end{figure}

The cell, with an internal path of 75 mm and containing purified
$^{85}$Rb with a small ($0.5$\%) admixture of $^{87}$Rb, no buffer
gas, and no wall coatings that might preserve polarization, is
modeled as a thermal equilibrium, Doppler-broadened vapor subject
to Zeeman shifts in the intermediate regime. The
atomic structure is calculated by diagonalization of the atomic
Hamiltonians $H_{\rm At}\supiso = H_{0}\supiso + H_{\rm
HFS}\supiso + H_{\rm Z}\supiso$, where $H_{0}\supiso$ is the
energy structure of the isotope $^{\rm iso}$Rb including fine-structure
contribution, $H_{\rm HFS}\supiso = g_{\rm HFS} \bJ \cdot \bI$ is
the hyperfine contribution, and $H_{\rm Z}\supiso = \bB \cdot(
g_{J} \bJ + g_{I} \bI )$ is the Zeeman contribution. All atomic parameters
are taken from references \cite{steckRb852010,steckRb872010}. The matrices $H_{\rm At}\supiso$ are numerically diagonalized to find field-dependent energy eigenstates, illustrated in Fig. \ref{fig:ED}, from which the complex linear
optical polarizability is calculated, including radiative damping. The complex refractive index $n_\pm$ for $\sigma_\pm$ polarizations is computed including Doppler broadening and a temperature-dependent atom density given by the vapor pressure times the isotope fraction, and the transfer function for the cell is calculated from the integral of the index along the beam path, including the measured drop in field strength of 15\% from the center to the faces of the cell.

\begin{figure}[t]
\centering
\includegraphics[width=0.45\textwidth]{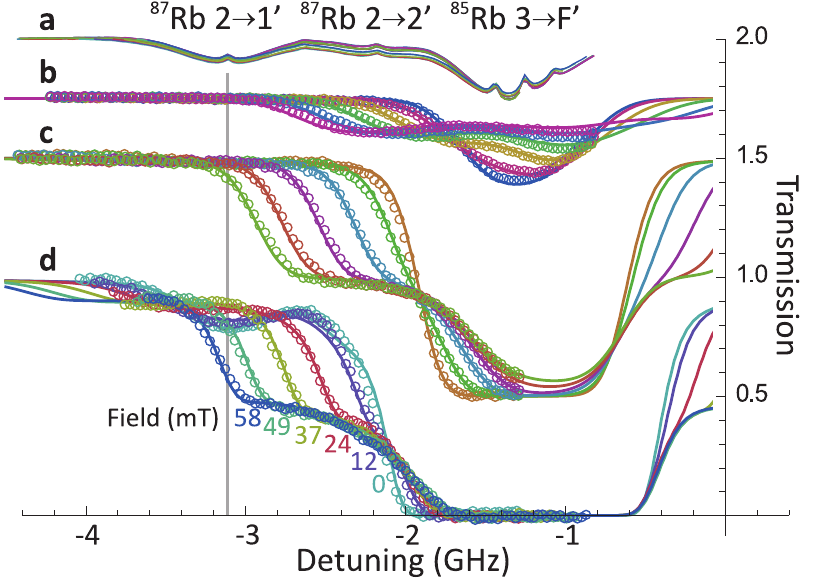}
\caption{Spectroscopic characterization of the rubidium atomic
ensemble. Circles show measured values, curves show predictions of
first-principles model (see text). {\bf a} Saturated-absorption
spectra acquired with a natural-abundance cell at room
temperature, as a frequency reference. Horizontal axis shows
detuning from the center of the $D_1$ spectral line. {\bf b}-{\bf
d} Transmission spectra for the cell containing $^{85}$Rb plus 0.5\% $^{87}$Rb at
temperatures of $22^\circ$C, $53^\circ$C and $83^\circ$C,
respectively. For each temperature, spectra with measured field
strengths (in mT) of 0, 12, 24, 37, 49, and 58 are shown, in order
of increasing line broadening. Grey vertical line shows $\omNOON$,
the probe detuning. This operating point gives strong Faraday
rotation with low absorption over the range 0-49 mT. Absorption
from the small residual $^{87}$Rb component can be seen
in {\bf d}. For clarity, parts {\bf a}-{\bf c} have been offset
vertically by 1, 0.75, and 0.5, respectively.
\label{fig:Spectroscopy}}
\end{figure}

\begin{figure}[t]
\centering
\includegraphics[width=0.4\textwidth]{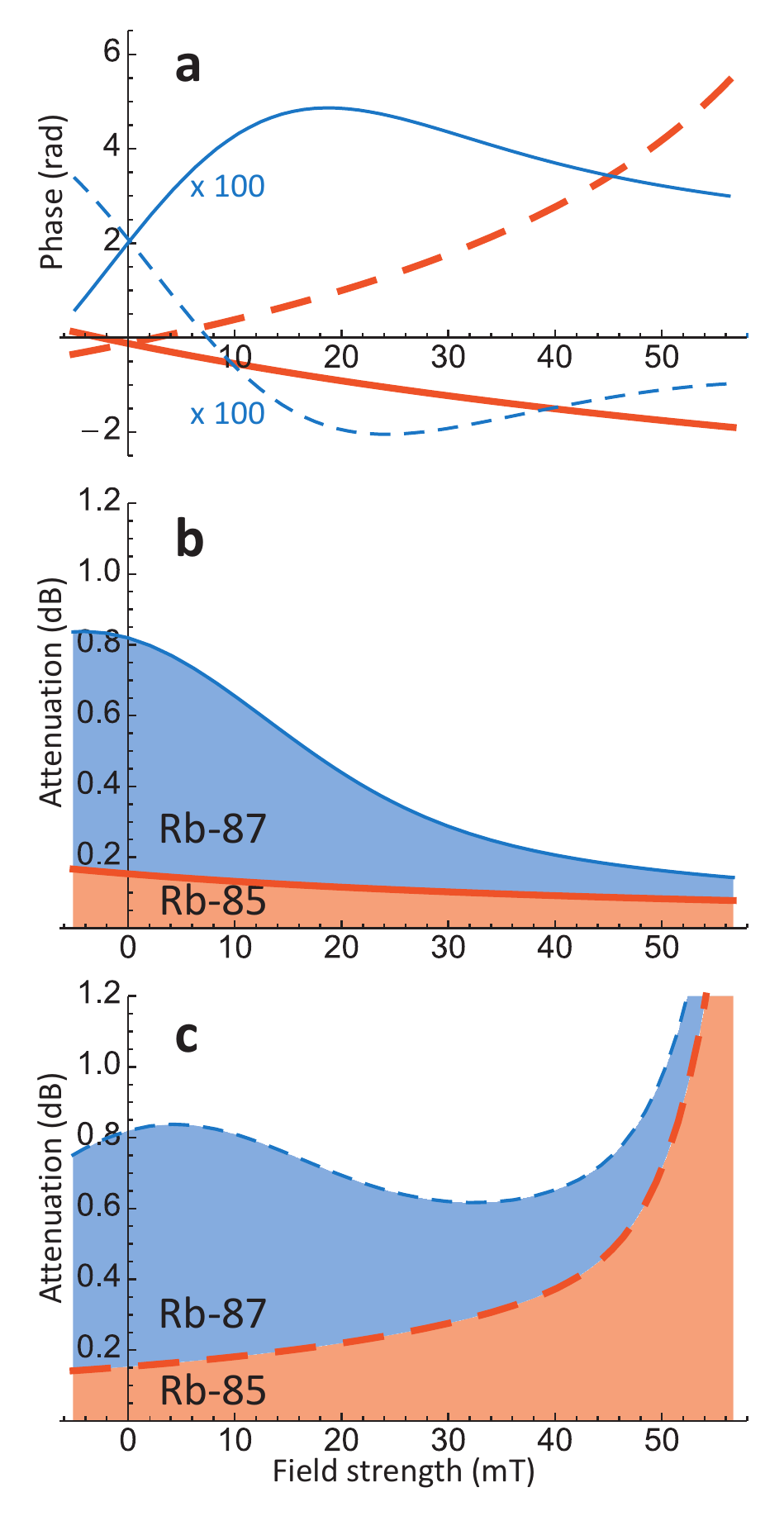}
\caption{Magneto-optical properties of the atomic ensemble, for
probe light at $\omNOON$, computed from model. Polarizations are
shown by solid ($\sigma_+$) and dashed ($\sigma_-$) curves.
Isotopic contributions are shown by thick orange ( $^{85}$Rb) and
thin blue ($^{87}$Rb) lines. {\bf a} Phase retardation vs. field
strength. The differential $\sigma_\pm$ retardation produces
optical rotation. Contribution of $^{85}$Rb is two orders of
magnitude larger than that of $^{87}$Rb, as expected from the
isotopic ratio, and in the opposite sense. {\bf b, c} Attenuation
vs. field strength for $\sigma_+,\sigma_-$ polarizations.
\label{fig:Transfer}}
\end{figure}

Transmission spectra, acquired with a low-power laser passing through the probe beam path, are used to validate the model, as shown in Fig. \ref{fig:Spectroscopy}. The frequency scale of the spectra is 
determined from simultaneous saturated absorption spectroscopy. Spectra taken at
temperatures $22^\circ$C, 53$^\circ$C, and 83$^\circ$C, and fields in the range 0-58 mT are used to determine the isotope fraction and to calibrate the field and temperature indicators, i.e., the current in the coils and the resistance of a thermistor near the cell, respectively. The transfer functions are shown in Fig. \ref{fig:Transfer}.

\subsection{Fisher information}
\begin{figure}[t]
\centering
\includegraphics[width=0.45\textwidth]{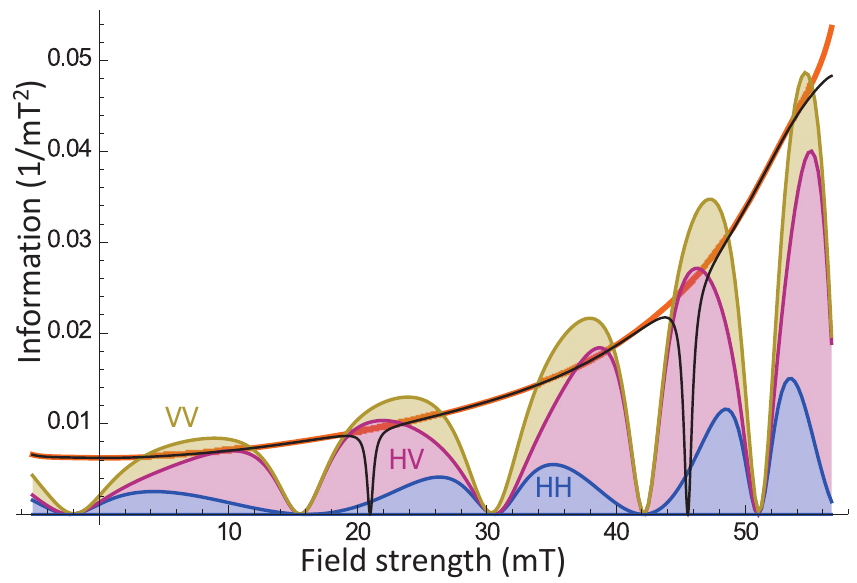}
\caption{Fisher information (FI) per input photon
in Faraday rotation probing of the ensemble. Thin black curve shows FI for
non-entangled photons of an arbitrarily chosen linear input polarization. Thick
orange curve shows the ``standard quantum limit,'' the largest FI
obtainable with non-entangled states. Solid regions indicate
contributions of $HH$, $HV$, and $VV$ outcomes to the \noon state FI,
calculated using $\rhoin$ from Fig.\ \ref{fig:DM_TC}. In all cases FI grows with increasing field, due to increased proximity to the $^{85}$Rb resonances.\label{fig:FIpip}
}
\end{figure}

The sensitivity of the measurement is computed using
the Fisher Information (FI) \cite{Braunstein1994SM}
\begin{equation}
{\cal I}(B) = \sum_i P_i (\partial_B \ln P_i)^2.
\label{eq:FIDef}
\end{equation}
Fig. \ref{fig:FIpip} shows ${\cal I}/2$, the FI per photon for the experimental \noon state $\rhoin$ shown 
in Fig. \ref{fig:DM_TC}. In an experiment with $N_{\rm tot} \gg 1$ total photons, the resulting magnetic uncertainty is $\delta B = (N_{\rm tot} {\cal I}/2)^{-1/2} $. Also shown is the standard quantum limit (SQL) for this figure of merit, i.e., the best possible FI per photon with non-entangled states, found by numerical optimization over possible single-photon inputs and the POVM $\{ \Pi_i \}$. The \noon state achieves 1.30$\pm$0.05 improvement over the SQL, at a field $B=37$ mT. All uncertainties given are one sigma statistical error. As with non-entangled photons (shown also in Fig. \ref{fig:FIpip}), the point of maximum sensitivity can be selected by varying the input polarization or by applying a bias magnetic field.

\subsection{Fisher information per damage}
The \noon state also gives an advantage in FI per scattered photon, the figure of merit for low-damage probing. We quantify scattering from the $^{85}$Rb ensemble as $\scat = {\rm Tr} [\rho \Pi_{\rm scat}]$, where $\Pi_{\rm scat} \equiv {\rm diag}(\lpp,\lpm,\lmp,\lmm)$ in the $\sigma_{\pm}$ basis and the mean number of scattering events is $s_{ab} \equiv 2-|t_a^{(85)}|^2-|t_b^{(85)}|^2$, where $t_\pm^{(85)}$ is the $^{85}$Rb contribution to $t_\pm$. A completely analogous calculation is made for single-photon scattering. As above, the the SQL for this figure of merit is found by numerical optimization. As shown in Fig. \ref{fig:FIpdMain}, the \noon state gives an advantage {$({\cal I}/\scat)_{\rm \noon}/({\cal I}/\scat)_{\rm SQL} = 1.40 \pm$ 0.06} over the SQL.

\subsection{State reconstruction and statistical error of Fisher information}
The state $\rho$ is found by quantum state tomography from the data shown in Fig. \ref{fig:SuperRes}, that is, by numerical search for the $\rho$ that minimises the chi-squared distance between the predictions of Eq. \ref{eq:CoincProbabilities} and the observed data. The optimized state is reported as the reconstructed $\rho$. To find statistical errors for the Fisher information, we search in the vicinity of $\rho$ for states $\rho'$ which maximize or minimize the Fisher information with a chi-squared distance at most $\delta$ larger than the optimum chi-squared. This establishes a statistical error which accounts for the non-linearity of the reconstruction procedure. The error ranges reported throughout are for $\delta = 1$, corresponding to $\pm 1 \sigma$ in the case of a simple linear reconstruction.

\subsection{Relation to state-of-the-art detection}
Thomas-Peter \emph{et al.} \cite{Thomas-Peter2011SM} consider the effect of detector inefficiency and derive visibility thresholds for metrological advantage in scenarios with constant loss. These thresholds are not directly applicable to our scenario with polarization-dependent, parameter-dependent losses. We note that intrinsic losses are included already in the FI calculation, and because they are field-dependent, provide some information about $B$, offsetting the loss of FI due to non-arrival of pairs. Constant extrinsic losses including detector inefficiency reduce the \noon FI by $\eta_{\rm ex}^2$ versus $\eta_{\rm ex}$ for any single-photon state. Current technology can achieve extrinsic efficiency of $\eta_{\rm ex} = \eta_{\rm det} \eta_{\rm path}$, with detector efficiency\cite{Lita2008SM} $\eta_{\rm det} = 0.95$ and source-to-detector path efficiency\cite{Vahlbruch2008SM} (including escape from the source cavity) $\eta_{\rm path} = 0.984$. With current technology, the \noon state demonstrated here gives a quantum enhancement of $1.21 \pm 0.05$ per sent photon and $1.15 \pm 0.04$ per photon scattered from the $^{85}$Rb ensemble ($1.31 \pm 0.06$ if the $^{87}$Rb contaminant is removed).

\end{document}